\documentclass[aps,12pt,showpacs,showkeys]{revtex4}

\usepackage[english]{babel}
\usepackage{amsmath}
\usepackage{amssymb}
\usepackage{amsbsy}
\usepackage{amstext}
\usepackage{graphicx}
\usepackage{pst-node}
\usepackage{verbatim}

\newcommand{\be}{\begin{eqnarray}}
\newcommand{\ee}{\end{eqnarray}}
\newcommand{\bdm}{\begin{displaymath}}
\newcommand{\edm}{\end{displaymath}}
\newcommand{\ds}{\displaystyle}
\newcommand{\ba}{\begin{array}}
\newcommand{\ea}{\end{array}}
\newcommand{\pa}[1]{\left(#1\right)}

\newcommand{\dpa}{\partial}

\begin{document}

\title{Gravitating binaries at the fifth post-Newtonian order in the post-Minkowskian approximation}

\author{Stefano Foffa}

\affiliation{D\'epartement de Physique Th\'eorique and Centre for Astroparticle Physics, Universit\'e de 
             Gen\`eve, CH-1211 Geneva, Switzerland}
\email{stefano.foffa@unige.ch}
\begin{abstract}
The Lagrangian governing the dynamics of a compact binary system is explicitly computed in the post-Minkowskian approximation, and at any post Newtonian order, by means of the
Effective Field Theory approach. 
This result is then specialized to the fifth post-Newtonian order and allows to determine one formerly unknown coefficient
of the energy expression for binary point masses on circular orbit as a function of the orbital angular frequency.

\end{abstract}

\keywords{classical general relativity, coalescing binaries, post-Newtonian 
expansion}

\pacs{04.20.-q,04.25.Nx,04.30.Db}

\maketitle

\section{Introduction}
The post-Newtonian (PN) approach to the 2-body problem in General Relativity
(see \cite{Blanchet_living} and \cite{Futamase} for two excellent reviews)  has been object of
a revived attention in recent years.

On one side, this is certainly related to the start of the era of gravitational wave
interferometers like LIGO and Virgo \cite{LIGOVirgo}, whose output data analysis requires a
very accurate knowledge of the binary dynamics, hence the necessity to carry on
calculations at very high order in the PN expansion parameter $v^2\sim G M/r$
\footnote{$v$ is the relative speed of the stars in units $c=1$,
and is related through the virial theorem to the curvature generated
by the whole binary system of total mass $M$}.

Besides this, the advent of the Effective Field Theory method developed in
\cite{Goldberger:2004jt}, \cite{EFTReview} (see also \cite{Foffa:2013qca} for a recent review)
opened the way to new computational possibilities at high PN orders,
while also triggering an healthy competition with more traditional approaches,
thus leading to the production of several new results both in the conservative \cite{PNcons}, \cite{Foffa:2012rn}, \cite{Jaranowski:2012eb},\cite{Jaranowski:2013lca}
and in the radiative \cite{PNrad} sector.

Another useful approach to the 2-body problem is the post-Minkowskian one,
namely an expansion in powers of $G$ only, discarding the link with $v$ expressed by the virial theorem.
Initially introduced to study the gravitational field generated by the binary system in the radiation zone 
\cite{postMink}, it can provide useful informations also to the conservative part of the binary dynamics.
A closed formula for the first post-Minkowskian (henceforth, simply denoted as PM) Hamiltonian,
and \emph{exact} in terms of the stars' momenta has been written in \cite{Ledvinka:2008tk} within the ADM formalism.
In the present work an analogous expression within the Lagrangian formalism is first derived via use of EFT methods
and then the 5PN part, containing terms of order $G v^{10}$, is explicitly extracted.
This allows to determine one formerly unknown coefficient of the 
energy-frequency relation for circular orbits at the fifth PN order.

\section{Main}
\label{main}
\subsection{Setup}
The derivation is done in the EFT framework along the lines of \cite{Foffa:2011ub},
which is here briefly reviewed in its basic aspects; the conventions of this paper for what concerns the metric signature, the definition of curvature tensors and other definitions are thus the same of \cite{Foffa:2011ub}.
The starting point is the action $S =S_{pp}+ S_{bulk}$, the first term being the worldline point particle action
\be
\label{az_EH}
S_{pp}=-\sum_{a=1,2} m_a\int {\rm d}\tau_a = 
-\sum_{a=1,2} m_a\int \sqrt{-g_{\mu\nu}(x^\mu_a) {\rm d}x_a^\mu {\rm d}x_a^\nu}\,,
\ee
while the second is the usual Einstein-Hilbert action
\footnote{
We adopt the ``mostly plus'' convention
$\eta_{\mu\nu}\equiv {\rm diag}(-,+,+,+)$, and the Riemann and Ricci tensors are
defined as $R^\mu_{\nu\rho\sigma}=\dpa_\rho\Gamma^\mu_{\nu\sigma}+
\Gamma^\mu_{\alpha\rho}\Gamma^\alpha_{\nu\sigma}-\rho\leftrightarrow\sigma$, 
$R_{\mu\nu}\equiv R^\alpha_{\mu\alpha\nu}$. } plus an harmonic gauge fixing term
\be
S_{bulk}=\frac{1}{16 \pi G}\int {\rm d}^4 x \sqrt{-g}\left[R-\frac{1}{2}\Gamma_\mu\Gamma^\mu\right]\,,
\ee
with $\Gamma^\mu\equiv \Gamma^\mu_{\alpha\beta}g^{\alpha\beta}$.
Notice that in the PM limit there are no divergent contributions in dimensional regularization,
so one can work in 4 space-time dimensions right from the beginning. 

Following \cite{Gilmore:2008gq},  the standard Kaluza-Klein (KK) 
parametrization  of the metric \cite{Kol:2007bc}, \cite{Kol:2010si} (a somehow similar 
parametrization was first applied within the framework of a PN calculation in
 \cite{annales}) is adopted here:
\be
\label{met_nr}
g_{\mu\nu}=e^{2\phi/m_p}\pa{
\ba{cc}
-1 & A_j/m_p \\
A_i/\Lambda &\quad e^{- 4\phi/m_p}\gamma_{ij}-
A_iA_j/m_p^2\\
\ea
}\,,
\ee
with $\gamma_{ij}=\delta_{ij}+\sigma_{ij}/m_p$, $m_p^{-2}=32 \pi G$, and the indices 
$i,j$ running over the $3$ spatial dimensions.
In terms of the metric parametrization (\ref{met_nr}),
the world-line coupling to the gravitational degrees of freedom
$\phi$, $A_i$, $\sigma_{ij}$  reads
\renewcommand{\arraystretch}{1.4}
\be
\label{matter_grav}
S_{pp}=- \sum_{a=1,2}m_a \ds \int {\rm d}\tau_a = \ds-m_a\int {\rm d}t_a\ e^{\phi/m_p}
\sqrt{\pa{1-\frac{A_i}{m_p}v_a^i}^2
-e^{-4 \phi/m_p}\pa{v_a^2+\frac{\sigma_{ij}}{m_p}v_a^iv_a^j}}\,,
\ee
\renewcommand{\arraystretch}{1.4}
and its Taylor expansion provides the following particle-gravity vertices:
\be
V_{\phi}(v_a)=-\frac{m_a}{m_p}\frac{1+v^2_a}{\sqrt{1-v^2_a}}\,,\quad V^i_A(v_a)=\frac{m_a}{m_p}\frac{v^i_a}{\sqrt{1-v^2_a}}\,,\quad V^{ij}_\sigma(v_a)=\frac{m_a}{m_p}\frac{v_a^iv_a^j}{2\sqrt{1-v^2_a}}\,.
\ee

The propagators for the gravitational degrees of freedom are needed as well. They can be derived by inverting the quadratic part of the pure gravity action $S_{bulk}$ written in terms of the KK variables:
\renewcommand{\arraystretch}{1.4}
\be
\label{bulk_action}
S_{bulk,free}&\subset&\frac{1}{32 \pi G}\int {\rm d}^4x
\left\{\frac{1}{4}\left[(\vec{\nabla}\sigma)^2-2(\vec{\nabla}\sigma_{ij})^2-\left(\dot{\sigma}^2-2(\dot{\sigma}_{ij})^2\right)\right]- 4 \left[(\vec{\nabla}\phi)^2-\dot{\phi}^2\right]\right.\nonumber\\
&&
+\left.\left[\frac{F_{ij}^2}{2}+\left(\vec{\nabla}\!\!\cdot\!\!\vec{A}\right)^2 -\dot{\vec{A}}^2 \right]\right\}\,.
\ee
\renewcommand{\arraystretch}{1.}
The propagators are fully relativistic and their spatial Fourier transforms are expanded in the quasi-static approximation as a sum of instantaneous contributions along the following pattern
\be
{\cal P}({\bf p}^2,t_a,t_b)=\frac{i}{{\bf p}^2-\partial_{t_a}\partial_{t_b}}\simeq
\frac{i}{{\bf p}^2}\left(1+\frac{\partial_{t_a}\partial_{t_b}}{{\bf p}^2}+
\frac{\partial_{t_a}^2\partial_{t_b}^2}{{\bf p}^4}\dots\right)\simeq
\frac{i}{{\bf p}^2}\left(1-\frac{\partial_{t_a}^2}{{\bf p}^2}+
\frac{\partial_{t_a}^4}{{\bf p}^4}\dots\right)
\ee
(the two expressions providing Lagragians terms that differ only by a total derivative)
up to the desired post-Newtonian order.

\subsection{The post-Minkowskian Lagrangian}
\begin{figure}
\includegraphics[width=1.\linewidth]{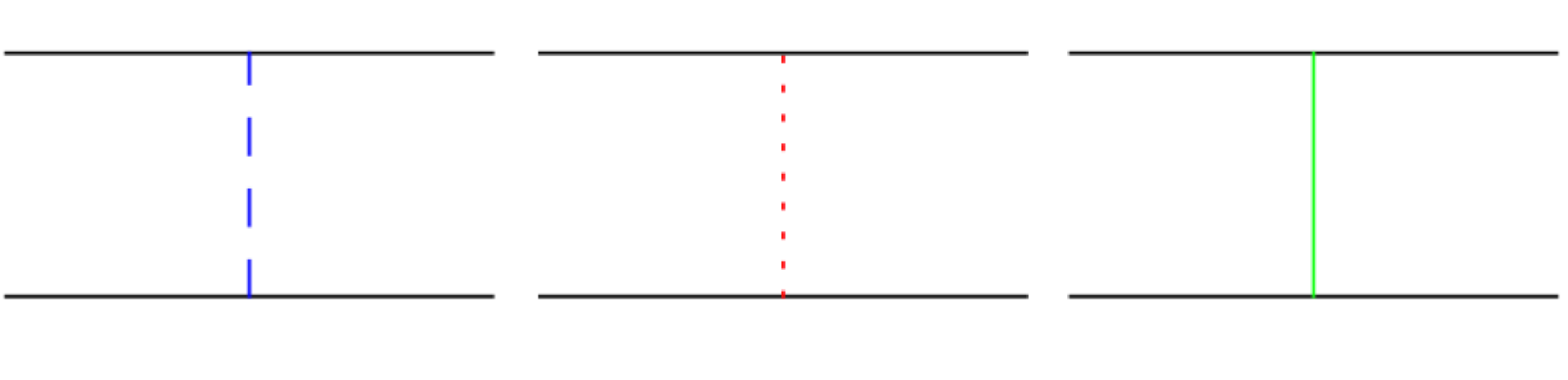}
\caption{The three diagrams giving the post-Minkowskian dynamics.  The $\phi$, $A$ and $\sigma$ 
      propagators are represented respectively by blue dashed, red dotted and 
      green solid lines.}
\label{diaG1}
\end{figure}

The relevant diagrams for the PM approximation are the three ones shown in figure \ref{diaG1}.
By means of the Feynman rules derived above, one finds amplitudes of this kind:
\be
A \sim G m_1 m_2 \int_{{\bf p},t_1,t_2}\sum_{n\geq0}\frac{(\partial_{t_1}\partial_{t_2})^n}{{\bf p}^{2 (n+1)}}\delta(t_1-t_2){\rm e}^{i {\bf p}.{\bf r}_{12}}V_k(v_1)V_k(v_2)\,,
\ee
where the measure of momentum integration is ${\rm d}^3 p/(2\pi)^3$, ${\bf r}_{12}\equiv \vec{x}_1(t_1)-\vec{x}_2(t_2)$, $v_{1,2}\equiv v_{1,2}(t_{1,2})$, and $k = \{\phi,A,\sigma\}$.

As momentum integration gives
\be\label{pint}
\int_{\bf p} \frac{1}{{\bf p}^{2 (n+1)}}{\rm e}^{i {\bf p}.{\bf r}_{12}}=\frac{\Gamma(3/2-n-1)}{\sqrt{4 \pi}^3\Gamma(n+1)} \left(\frac{r_{12}}{2}\right)^{2 n-1}\,,
\ee
the complete PM potential (at all post-Newtonian orders) can be written in the following way:
\be\label{VPM}
V_{\rm PM}&=&-G m_1 m_2 \sum_{n\geq0} \frac{\Gamma(3/2-n-1)}{4^n\sqrt{\pi} \Gamma(n+1)}\left\{(\partial_{t_1}\partial_{t_2})^n\left[r_{12}^{2n-1}f(v_1,v_2)\right]\right\}_{t_1=t_2}\,,\\
\label{fofv}
&&f(v_1,v_2)\equiv\frac{1+ v_1^2+v_2^2-4\vec{v_1}\cdot\vec{v_2}-v_1^2v_2^2+2(\vec{v_1}\cdot\vec{v_2})^2}{\sqrt{(1-v_1^2)(1-v_2^2)}}
\ee

The above equation, which contains higher-than-second derivatives, can be manipulated by adding suitable total derivative terms and/or using the double zero trick \cite{multizero},
that transforms terms quadratic in the accelerations into ${\cal O}(G^2)$ ones. The result is a Lagrangian which is classically equivalent to the above expression, but which is just linear in the accelerations and does not contain higher derivatives.
Such possibility will be exploited in the next subsection to determine the binary dynamics at 5PN in the post-Minkowskian approximation in the center-of-mass frame.

Before doing so, it is worth noticing that one can eliminate also the residual accelerations\footnote{To be more precise, accelerations are traded for their equations of motion, which for the PM point of view means eliminating them as they will give rise to ${\cal O}(G^2)$ terms.} in the Lagrangian by means of a series of contact transformations \cite{Damour:1990jh} bringing from $(\vec{r},\vec{v}_{1,2})$ to some new variables $(\vec{R},\vec{V}_{1,2})$, such that one is left with an acceleration-free generalized potential $V^*_{PM}(\vec{R},\vec{V}_{1,2})$ which can assume this remarkably simple form:
\be\label{V*simple}
V^*_{PM}=-\frac{G m_1 m_2}{R}\frac{f(V_1,V_2)}{\sqrt{1-V_1^2+\left(\frac{\vec{R}\cdot\vec{V}_1}{R}\right)^2}}\,.
\ee
It has to be said that the above expression is not unique (that is, there are infinite contact transformations that eliminate accelerations), and that a price has to be paid: the contact transformations generally do not respect the harmonic gauge conditions; more than that, the new gauge condition has been proven to explicitly break Lorentz invariance \cite{Martin:1979ph}.

Along with the kinetic term $m_a\sqrt{1-v_a^2}$, equations (\ref{VPM}) or (\ref{V*simple}) make the Lagrangian counterpart of the expression derived in \cite{Ledvinka:2008tk} within the Hamiltonian formalism. To prove that the two formalisms are indeed equivalent, one should show explicitly that they are related by a Legendre transformation, eventually after a further contact transformation (or canonical transformation, on the Hamiltonian side). An alternative way (actually more viable at high PN's) is to show directly that the two formalisms provide the same predictions for gauge invariant quantities, like the Energy of circular orbits as a function of the orbital frequency (measured at spatial infinity); this method has indeed been used to prove that \cite{Foffa:2012rn} and \cite{Jaranowski:2012eb} are two equivalent descriptions of the binary dynamics at 4PN in the post-post-Minkoskian approximation.
The next subsection will be thus devoted to work out the prediction of the PM Lagrangian formalism for the energy of circular orbits at 5PN.

\subsection{Dynamics at 5PN}

As previously mentioned, eq.(\ref{VPM}) can be transformed, within the harmonic gauge, into in expression which is linear in the accelerations.
Such Lagrangian (whose expression is rather long and not worth reporting) can be used, along the lines described in \cite{Lorentz}, to determine the center of mass position at the needed order, 5PN in this case \footnote{The 4PN order in the center of mass position is sufficient for the purposes of the present work;
however the derivation at 5PN in the PM limit is a very useful test of Lorentz invariance, and thus of the correctness of the calculation.}, as well as the energy in a generic frame.
Then, the knowledge of the center of mass position is exploited (along with the equations of motion) to write the energy in the center of mass frame, as a function of the relative position $r$ and velocity $v$, as well as of the reduced mass $\mu$ and of the symmetric mass ratio of the system $\nu\equiv \mu/M$:
\be\label{Ecm}
E^{\rm PM}_{\rm 5PN}=&&\frac{\mu v^{12}}{1024}\left(231-5401 \nu+49632 \nu^2-224675 \nu^3+502425 \nu^4-445005 \nu^5\right)\nonumber\\
+&&\frac{G M \mu}{256 r}\left[v^{10}\left(1197-14505\nu+64818\nu^2-103695 \nu^3-41377\nu^4+195615\nu^5\right)\right.\nonumber\\
-&& v^8 v_r^2 \nu \left(2541-17233\nu+12815\nu^2+125626\nu^3-232875\nu^4\right)\nonumber\\
+&& 2 v^6 v_r^4 \nu \left(495-1410\nu-12471\nu^2+58959\nu^3-66825\nu^4\right)\nonumber\\
+&& 10 v^4 v_r^6 \nu \left(11-410\nu+2701\nu^2-6062\nu^3+4131\nu^4\right)\nonumber\\
-&& 35 v^2 v_r^8 \nu \left(7-80\nu+313\nu^2-473\nu^3+207\nu^4\right)\nonumber\\
+&&\left. 9v_r^{10} \nu \left(7-63\nu+189\nu^2-210\nu^3+63\nu^4\right)\right]+{\cal O}(G^2)\,,
\ee
with $v_r\equiv \vec{v}\cdot\vec{r}/r$.
Neglecting gravitational wave radiation, the PM Energy is conserved under the action of the PM equation of motion, which has the following form in the center of mass frame:
\be\label{acm}
\vec{a}&=&\vec{a}_{\rm 0-3PN}+\vec{a}^{\rm PM}_{\rm 4PN}+\vec{a}^{\rm PM}_{\rm 5PN}+{\cal O}(6PN)\,,\nonumber\\
\vec{a}^{\rm PM}_{\rm 4PN}&=& \frac{G \mu}{256 r^3}\left\{2\left[16 v^8\left( -21+175\nu-488\nu^2+432\nu^3\right)+24 v^6 v_r^2\left( 56-397\nu+872\nu^2-528\nu^3\right)\right.\right.\nonumber\\
&&-120 v^4 v_r^4\left(18-125\nu+269\nu^2-178\nu^3\right)+280v^2 v_r^6\left(5-34 \nu+69\nu^2-42\nu^3\right)\nonumber\\
&&-\left.315 v_r^8\left(1-7 \nu+14\nu^2-7\nu^3\right)\right] \vec{r}\nonumber\\
&&+\left[16 v^6\left(157-864\nu+1104\nu^2+384\nu^3\right)-48 v^4 v_r^2\left(120-623\nu+688\nu^2+288\nu^3\right)\right.\nonumber\\
&&+\left.\left. 240v^2 v_r^4\left(22-111\nu+115\nu^2+28\nu^3\right)-560 v_r^6\left(3-14\nu+11\nu^2+2\nu^3\right)\right]v^r \vec{v}\right\}\,,\nonumber\\
\vec{a}^{\rm PM}_{\rm 5PN}&=& \frac{G \mu}{256 r^3}\left\{\left[4 v^{10}\left( -163+2093\nu-10124\nu^2+21360\nu^3-16192\nu^4\right)\right.\right.\nonumber\\
&&+6 v^8 v_r^2\left(576-6163\nu+24084\nu^2-39184\nu^3+20416\nu^4\right)\nonumber\\
&&-120 v^6 v_r^4\left(64-635\nu+2287\nu^2-3509\nu^3+1896\nu^4\right)\nonumber\\
&&+140v^4 v_r^6\left(56-521 \nu+1713\nu^2-2345\nu^3+1120\nu^4\right)\nonumber\\
&&-\left.630 v^2 v_r^8\left(6-54 \nu+167\nu^2-205\nu^3+79\nu^4\right)+693v_r^{10}\left(1-9\nu+27\nu^2-30\nu^3+9\nu^4\right)\right] \vec{r}\nonumber\\
&&+2\left[v^8\left(1493-13616\nu+41376\nu^2-37120\nu^3-14080\nu^4\right)\right.\nonumber\\
&&-12 v^6 v_r^2\left(384-3209\nu+8660\nu^2-6536\nu^3-2816\nu^4\right)\nonumber\\
&&+30v^4 v_r^4\left(208-1621\nu+3953\nu^2-2712\nu^3-672\nu^4\right)\nonumber\\
&&-\left.\left.140v^2 v_r^6\left(28-201\nu+421\nu^2-213\nu^3-40\nu^4\right)+315 v_r^8\left(3-20\nu+36\nu^2-12\nu^3-2\nu^4\right)\right]v^r \vec{v}\right\},\nonumber\\
\ee
where $\vec{a}_{\rm 0-3PN}$ can be read off \cite{Blanchet_living}, section 9.3,  $\vec{a}_{\rm 4PN}^{\rm PM}$ can be deduced from the results of \cite{Foffa:2012rn} and \cite{Foffa:2013qca}, and finally
$\vec{a}_{\rm 5PN}^{\rm PM}$ has been determined from the same Lagrangian that led to eq.(\ref{Ecm}).

Equations (\ref{Ecm}) and (\ref{acm}) (together with the known results at lower PN's) are sufficient to determine the $\nu^5$ coefficient of the energy-frequency relation for circular orbits, where the frequency has been traded of the adimensional quantity
$x\equiv(G M \omega_{circ})^{2/3}$. Indeed, ${\cal O}(G^2)$ energy terms, even if generally dependent on $\nu^5$ in this gauge, do not give contribution to the energy-frequency relation at this PN order, as it has been explicitly checked for any possible
${\cal O}(G^2 v^8)$ Lagrangian structure compatible with Lorentz invariance: this fact is more transparent in the Hamiltonian approach, where it is clear that at the $n^{th}$ PN order  the $\nu^n$ terms of the center-of-mass Hamiltonian can have at most one power of $G$,
see for instance \cite{Jaranowski:2013lca}.

A tedious but straightforward calculation finally leads to the determination of the $\nu^5$ coefficient of the energy frequency relation for circular orbits at 5PN. This has to be added to other previously known terms to give:
\be\label{Ecirc}
E(x)_{5PN}^{circ}&=&-\frac{\mu x^6}{2}\left[-\frac{45927}{512}-\left(55.13(3)+\frac{4988}{35}\log{x}\right)\nu\right.\nonumber\\
&& \left.+\left(c_2 -\frac{656}{5}\log{x}\right)\nu^2+c_3 \nu^3+c_4 \nu^4+ \frac{3121}{32}\nu^5\right]\,,
\ee
where the first coefficient, corresponding to the limit $\nu\rightarrow0$  is easily derivable from the test particle dynamics in the Schwarzschild background, while the other formerly known terms have been computed in \cite{LeTiec:2011ab},
except for the logarithmic term in the $\nu$ coefficient which has been derived in \cite{Blanchet:2010zd}.

\section{Conclusions}
The 5PN dynamics is not of immediate interest for the upcoming gravitational wave detection and phenomenology, but it may become relevant for the next generation of detectors.
It is also worth reminding that the effacement principle ceases its effect precisely at this order for spin-less stars, thus making the 5PN dynamics very relevant for possibly getting informations about the internal structure of compact objects (see \cite{Yagi:2013baa,Favata:2013rwa} for two very recent works in this direction).
As can be seen from eq.(\ref{Ecirc}), the test-particle limit and the post-Minkowskian approximation attack from opposite sides the energy-frequency relation; as already happened in the 4PN case, a good roadmap to determine the remaining unknown coefficients
(which depend on the $2{nd}$ Newtonian Love number of the compact star) is to work using self-force methods on one side (low $\nu$), and ADM or EFT  approaches on the other one (terms with high powers of $\nu$).
\label{se:conclusion}

\section*{Acknowledgements}
I would like to thank the referee for making some pertinent observations.
This work is supported by the Fonds National Suisse.

\section*{Appendix: derivation of $V^*_{PM}$}
Since all the Lagrangian terms containing accelerations can be removed by means of the double zero trick and by a suitable contact transformation,
one can obtain $V^*_{PM}$ simply by setting to zero all such terms that are generated by doing the derivatives in equation  (\ref{VPM}), this implying also that one can completely
avoid to apply the derivatives to the function $f(v_1,v_2)$.

Actually it is more convenient to take a step backwards, and apply the operator $(-1)^n(\partial_{t_1})^{2n}$ (which is equivalent to $(\partial_{t_1}\partial_{t_2})^n$ modulo a derivation by parts in the final result\footnote{The use of either of the two equivalent operators for some $n$ brings to different expression for $V^*_{PM}$}.) directly in eq.(\ref{pint}); again neglecting accelerations one has:
\be
&&(-1)^n\partial_{t_1}^{2n} \int_{\bf p} \frac{1}{{\bf p}^{2 (n+1)}}{\rm e}^{i {\bf p}.{\bf R_{12}}}\simeq V_1^{i_1}\dots V_1^{i_{2n}}I^{i_1\dots i_{2n}}\,,\\
&&I^{i_1\dots i_{2n}}\equiv\int_{\bf p} \frac{p^{i_1}\dots p^{i_{2n}}}{{\bf p}^{2 (n+1)}}{\rm e}^{i {\bf p}.{\bf R}_{12}}\,.
\ee
The value of the integral can be deduced from (\ref{pint}) by taking gradients with respect to $\vec{R}_{12}$ and the result can be written as
\be
I^{i_1\dots i_{2n}}=\frac{1}{2^{2n-1}}\frac{(-1)^n\Gamma\left(\frac32-n-1\right)}{(4 \pi)^{3/2} \Gamma\left(n+1\right)}\sum_{j\leq n}\left\{\eta^{ab}\right\}^{n-j}\{R^c\}^{2j} P^{n+j}\left[R^{2n-1}\right]\,,
\ee
where $\left\{\eta^{ab}\right\}^{n-j}$ and $\{R^c\}^{2j}$ indicate the product of $(n-j)$ metric tensors (symmetrized over the indices) and of $2j$ three-vectors $\vec{R}$, respectively, while the operator $P[f]$ is defined as
\be
P[f]\equiv\frac{1}{R}\frac{{\rm d} f}{{\rm d} R}\Rightarrow P^{n+j}\left[R^{2n-1}\right]=(-1)^{j+1}(2n-1)!!(2j-1)!!\frac{1}{R^{2j+1}}\,.
\ee
The next step of the derivation involves combinatorics as, for given $n$ and $j$, one can contract $2j$ $\vec{V}_1$ vectors with $\{R^c\}^{2j}$ in $B(2n,2j)$ ways ($B(a,b)$ denoting the binomial coefficient), while the remaining $2(n-j)$ $\vec{V}_1$'s contract with the $\left\{\eta^{ab}\right\}^{n-j}$ tensor in $(2n-2j-1)!!$ ways.
One is thus led to
\be\label{Vnoacc}
V^*_{PM}&=&\frac{G m_1 m_2}{R}\sum_{\{n,j\}}C_{nj}\left(\frac{\vec{R}\cdot\vec{V}_1}{R}\right)^{2j}{V_1^2}^{(n-j)}f(V_1,V_2)\,,\\
C_{nj}&=&\frac{(-1)^{n+j+1}}{4^n\sqrt{\pi}}\frac{\Gamma\left(\frac{3}{2}-n-1\right)}{\Gamma\left(n+1\right)}\frac{(2n-1)!! (2j-1)!!(2n)!(2n-2j-1)!!}{(2j)!(2n-2j)!}\label{Cnjwz}\,.
\ee
Finally the following resummation formula
\be
\sum_{\{nj\}}C_{nj}x^{2j}y^{2(n-j)}=-\frac{1}{\sqrt{1+x^2-y^2}}
\ee
brings to to eq. (\ref{V*simple}), which can be eventually symmetrized:
\be
V^*_{PM}=-\frac{G m_1 m_2}{2R}\left[\frac{1}{\sqrt{1-V_1^2+\left(\frac{\vec{R}\cdot\vec{V}_1}{R}\right)^2}}+\frac{1}{\sqrt{1-V_2^2+\left(\frac{\vec{R}\cdot\vec{V}_2}{R}\right)^2}}\right]f(V_1,V_2)\,.
\ee

\end{document}